\documentclass[onecolumn]{aa}
\usepackage{graphicx}
\usepackage{txfonts}
\begin{document}
\title{ Transport and mixing in the radiation zones of rotating stars: I-Hydrodynamical processes}
\titlerunning{Transport and mixing in rotating stars: Hydrodynamical processes}  
\authorrunning{Mathis \& Zahn}  

  \author{S. Mathis                
	 \and              
	  J.-P. Zahn             
	  }

   \offprints{J.-P. Zahn}

   \institute{LUTH, Observatoire de Paris, F-92195 Meudon\\
              \email{stephane.mathis@obspm.fr; jean-paul.zahn@obspm.fr}     
            }

   \date{Received February 17; accepted June 3, 2004}

  \abstract{
 The purpose of this paper is to improve the modelization of the rotational mixing which occurs in stellar radiation zones, through the combined action of the thermally driven meridional circulation and of the turbulence generated by the shear of differential rotation. The turbulence is assumed to be anisotropic, due to the stratification, with stronger transport in the horizontal directions than in the vertical. The main difference with the former treatments by Zahn (1992) and Maeder \& Zahn (1998) is that we expand here the departures from spherical symmetry  to higher order, and include explicitly  the differential rotation in latitude, to first order. This allows us to treat simultaneously the bulk of a radiation zone and its tachocline(s). 
  Moreover, we take fully into account  the non-stationarity of the problem, which will enable us
  to tackle the rapid phases of evolution.      
  The system of partial differential equations, which govern the transport of angular momentum, heat and chemical elements, is written in a form which makes it ready to implement in a stellar evolution code. Here the effect of a magnetic field is deliberately ignored; it will be included in forthcoming papers.
  
   }

   \maketitle
%

  \section{Introduction - Motivation}
In rotating stars, radiation zones are generally not in radiative equilibrium, as was shown by Von Zeipel (1924). Eddington (1925) and Vogt (1925) concluded that this thermal imbalance drives a large-scale meridional circulation, which was described in detail by Sweet (1950) for a given rotation law. Mestel (1953) discussed the role of composition inhomogeneities, which build up through the nuclear burning of hydrogen; he showed that they would slow down the circulation, and possibly prevent any mixing. Partly for this reason, rotational mixing was largely ignored in further work on stellar evolution. The next significant step was made by Busse (1982) who showed that  angular momentum would be advected in such a way as to achieve a state of zero circulation, at least in an non-evolving, inviscid star with no stress applied to its surface. 

The subject has been re-examined in Zahn (1992), who assumed that hydrodynamical instabilities due to the shear of differential rotation would enforce a rotation law with much larger variation in the vertical than in the horizontal direction. With such `shellular' rotation, the problem reduces to one dimension (or rather 1.5 dimension, since departures from spherical symmetry are taken in account), and this renders  it much more tractable. In particular one can then treat consistently the redistribution of angular momentum through the meridional circulation, and thus its feedback on the rotation profile, while preserving its advective nature.  This makes it possible to prove that the boundary condition applied at the surface on the angular momentum flux plays a crucial role. In a star which loses angular momentum through a wind, the meridional circulation is enforced by the requirement of transporting this momentum to the surface. On the contrary, when the star loses no angular momentum, it relaxes to a state where the advection of momentum is balanced by its transport through turbulent motions. Thereafter this treatment was extended to inhomogeneous stars, with a general equation of state, and partly adapted to phases of fast stellar evolution (Maeder \& Zahn 1998).  

The first evolutionary sequence calculated according to  these prescriptions was that of a 9 M$_\odot$ star (Talon et al. 1997); it was followed by extensive grids of models built by Meynet and Maeder (2000), who refined the prescription for the loss of angular momentum by taking into account the latitudinal variation of the wind. For massive stars, these models are in better agreement with observational data than models without rotational mixing: they account for the observed surface composition, and they predict the correct ratio between red and blue supergiants observed in open clusters and in the Magellanic clouds (Maeder \& Meynet 2001). They also explain the absence of light elements depletion on the blue side of the Li dip, as shown by Talon and Charbonnel (1998).
But they fail to correctly reproduce the flat rotation profile in the radiative interior of solar-type stars (Mathias \& Zahn 1997). In such stars, which are slow rotators, other physical processes must therefore contribute to the transport of angular momentum; the most plausible candidates are either magnetic stresses, as advocated already by Mestel (1953), or internal gravity waves emitted at the interface with convection zones (Talon et al. 2002; Talon \& Charbonnel 2003).
\medskip

The purpose of the present paper is to improve the modeling of rotational mixing, in particular during the phases of fast stellar evolution, when steep vertical gradients develop, both of angular velocity and  of chemical composition. To achieve this, our aim is threefold :
\begin{itemize}
\vskip -3pt
\item we extend the linear expansion of all variables to higher order spherical harmonics, in order to capture the tachocline circulation which is induced in its vicinity by the differential rotation of a convection zone,
\item for the same reason, we treat explicitly the departure from shellular rotation, in the linear approximation,
\item we retain all time derivatives, except those describing the relaxation to hydrostatic and geostrophic balance, in order to better resolve the phases of fast stellar evolution.
\end{itemize}

In forthcoming papers, we shall introduce a magnetic field, and examine its impact on the thermally driven circulation.
  
  \section{Main assumptions}
Our treatment is based on the
assumption that the rotating star is only weakly two-dimensional, and this for two reasons. The first is that the rotation rate is sufficiently moderate to allow the centrifugal force to be considered as a perturbation, compared to the gravitational field. The second reason rests on the conjecture that the differential rotation induced by the meridional circulation gives rise to turbulent motions which are strongly anisotropic due to the stable stratification, with much stronger transport in the horizontal directions than in the vertical: i.e. $ \nu_{v} \ll  \nu_{h}$  and $D_{v} \ll D_{h}$
 respectively for the turbulent viscosity and diffusivity. Such anisotropic turbulence is observed in the Earth's atmosphere and oceans; in a star we expect it to smooth the horizontal variations of angular velocity and of chemical composition, a property we shall invoke to discard certain non-linear terms.
 The prescriptions to be used for these turbulent diffusivities are discussed and updated in a companion paper (Mathis et al. 2004).

  We thus consider an axisymmetric star, and 
assume that the horizontal variations of all quantities are small and smooth enough to allow their linearization and their expansion in a modest number of spherical harmonics $P_{l}(\cos\theta)$. As reference surface, we chose either the sphere or the isobar, and write all scalar quantities either as
  \begin{equation}
  X(r,\theta)=X_{0}(r)+\sum_{l}\widehat{X}_{l}(r)P_{l}\left(\cos\theta\right)
  \end{equation}
  or
  \begin{equation}
  X(P,\theta)=\overline{X}(P)+\sum_{l>0}\widetilde{X}_{l}(P)P_{l}\left(\cos\theta\right) .
  \end{equation}
Let us establish the relation between those two modal expansions. We expand the pressure around the sphere as:
  \begin{equation}
  P(r,\theta)=P_{0}(r)+\sum_{l>0}\widehat{P}_{l}(r)P_{l}\left(\cos\theta\right) ,
  \end{equation}
 and introduce the radial coordinate of the isobar
   \begin{equation}
  r_P(r, \theta) = r + \sum_{l>0}\xi_{l}(r) P_{l}(\cos\theta) ,
  \end{equation}
  where $r$ is the mean value of the radius of an isobar. Taking the Taylor expansion of $P$ to first order, we have:  
  \begin{equation}  P\left(r+\sum_{l>0}\xi_{l}(r)P_{l}(\cos\theta),\theta\right)=P_{0}(r)+\sum_{l>0}\widehat{P}_{l}(r)P_{l}(\cos\theta)+ \left( \frac{ {\rm d} P_{0}}{ {{\rm d} r}}\right)   \sum_{l>0}\xi_{l}(r)P_{l}(\cos\theta) ,
  \end{equation}
and since by definition the pressure is constant on the isobar, we conclude that  
  \begin{equation}
  \xi_{l}(r)=-\frac{\widehat{P}_{l}(r)} {{{\rm d} P_{0}/{\rm d}r}} .
  \end{equation}
  
If we apply the same procedure to any other variable $X$, we get   
     \begin{equation}
X\left(r+\sum_{l>0} \xi_{l}(r)P_{l}(\cos\theta), \theta\right)=X_{0}(r)+\sum_{l>0}\left[\widehat{X}_{l}(r)-\left(\frac{{\rm d}X_{0}}{{\rm d}P_{0}}\right)\widehat{P}_{l}(r)\right]P_{l}(\cos\theta)
  \end{equation}
and therefore 
  \begin{equation}
  \tilde{X}_{l}({r})=\widehat{X}_{l}(r)-
     \left( \frac{{\rm d} X_{0}}  {{\rm d} P_{0}} \right)  \widehat{P}_{l}(r) .
     \label{xtilde}
  \end{equation}
 This relation will serve below in \S5.1 to calculate the effective gravity.

  \section{Transport of angular momentum}
  
 We start from the momentum equation
     \begin{equation}
  \rho\left[ \partial_{t}\vec V+\left( \vec V \cdot \vec \nabla \right) \vec V \right]=-\vec\nabla P - \rho\vec\nabla\phi+\vec\nabla\cdot ||\tau|| \label{NV-vec}
  \end{equation} 
where $\rho$ is the density, $\phi$ the gravitational potential, and $||\tau||$ represents the turbulent stresses. The macroscopic velocity field  $\vec V$ is the sum of a zonal flow with angular velocity $\Omega(r,\theta)$ and a meridional flow $\vec{\mathcal{U}}(r, \theta)$:
  \begin{equation}
 \vec V=r\sin\theta \, \Omega(r,\theta) \, {\vec e}_{\varphi}+\vec{\mathcal{U}}(r, \theta) .
\end{equation}
  The latter can be split into a spherically symmetric part, which represents the contraction or dilation of the star during its evolution, plus the meridional circulation
\begin{equation}
  \vec{\mathcal{U}}=\dot{r} \, {\vec e}_{r}+\vec{\mathcal{U}}_{M}(r, \theta) ,
  \end{equation}  
which we expand in spherical functions
  \begin{equation}
\vec{\mathcal{U}}_{M}=\sum_{l>0}\left[U_{l}(r)P_{l}\left(\cos\theta\right){\vec e}_{r}+V_{l}(r)\frac{{\rm d}P_{l}(\cos\theta)}{{\rm d}\theta}{\vec e}_{\theta}\right] .
\label{merid-exp}
  \end{equation}
 Using the continuity equation in the anelastic approximation, i.e. $\vec{\nabla} \cdot (\rho\,\vec {\mathcal{U}}_{M})=0$, we obtain the following relation between $U_{l}(r)$ and $V_{l}(r)$:
  \begin{equation}
  V_{l}=\frac{1}{l(l+1)\rho r}{{\rm d} \over {\rm d}r} \left(\rho r^2U_{l}\right) .
  \end{equation}   
Making again use of the continuity equation, the azimuthal component of (\ref{NV-vec}) can be written as an advection/diffusion equation for the angular momentum:
   \begin{equation}
\rho {{\rm d} \over {\rm d}t} \left(r^{2} \sin^{2} \theta \, \Omega \right) + \vec \nabla \cdot \left( \rho r^{2} \sin^{2} \theta \, \Omega \, \vec{\mathcal{U}_M} \right) 
= \frac{\sin^{2}\theta}{r^2} \partial_{r} \left( \rho \nu_{v}r^{4} \partial_{r} \Omega \right) + \frac{1}{\sin\theta} \partial_{\theta} \left( \rho\nu_{h} \sin^{3} \theta \, \partial_{\theta} \Omega \right) ,
\label{AM-diff-adv}
\end{equation} 
  where as in Zahn (1992) we assume that the effect of the turbulent stresses on the large scale flow are adequately described by an anisotropic eddy viscosity, whose components are $\nu_{v}$ and $\nu_{h}$ respectively in the vertical and horizontal directions. 
 As discussed in Mathis et al. (2004), they act to reduce the cause of turbulence, as observed in laboratory experiments of rotating flows, namely here the vertical and horizontal gradients of angular velocity. This explains why the respective fluxes of angular momentum contain only these gradients, in contrast with the treatment of Kippenhahn (1963), who considered the effect of an imposed anisotropic turbulence, due to thermal convection; in that case the rotation becomes non-uniform, as is actually observed in the solar convection zone. 
Note the introduction of the lagrangian time-derivative ${{\rm d} / {\rm d}t}$, meaning that   the radial coordinate $r$ is the mean radius of the layer (either sphere or isobar) which encloses the mass $M_{r}$, with ${\rm d}M_{r}=4\pi\rho r^{2}{\rm d}r$. 
  
The form of eq.  (\ref{AM-diff-adv}) incites to expand the angular velocity as
 \begin{equation}
  \Omega(r,\theta)=\overline{\Omega}(r)+ \widehat{\Omega}(r,\theta) 
 =  \overline{\Omega}(r)+ \sum_{l>0}\Omega_{l}(r) \, Q_l(\theta) ,
 \label{omega-expand}
  \end{equation}
 with the horizontal average being defined as
  \begin{equation}
  \overline{\Omega}(r)=\frac{\int_{0}^{\pi}\Omega(r,\theta)\sin^{3}\theta \, {\rm d} \theta}{\int_{0}^{\pi}\sin^{3}\theta  \, {\rm d} \theta} ,
   \end{equation}
 and where the horizontal functions $Q_l(\theta)$ satisfy the orthogonality condition
 \begin{equation}
\frac{\int_{0}^{\pi} Q_l(\theta) \sin^{3}\theta \, {\rm d} \theta}
{\int_{0}^{\pi}\sin^{3}\theta  \, {\rm d} \theta} =0.
  \end{equation}
  These horizontal functions are readily identified:
 \begin{equation}
  Q_l(\theta)=  P_{l}\left(\cos\theta\right) - I_l   \quad  \hbox{with} \quad 
  I_l =    \frac{\int_{0}^{\pi} P_l(\cos\theta) \sin^{3}\theta \, {\rm d} \theta}
{\int_{0}^{\pi}\sin^{3}\theta  \, {\rm d} \theta}  =    \delta_{l,0} - {1 \over 5} \delta_{l,2} \, ;
  \label{fal}
   \end{equation}
except for $l=2$, these functions $Q_l$ reduce to the Legendre polynomials.

  \subsection{Vertical transport of angular momentum}
  
  Taking the horizontal average of equation (\ref{AM-diff-adv}), and using the assumption that $\overline{\Omega} \gg \Omega_{l}$, we obtain the following vertical advection/diffusion equation for the mean angular velocity $\overline{\Omega}$:  
    \begin{equation}
  \rho {{\rm d} \over {\rm d}t} (r^2\overline{\Omega})=\frac{1}{5r^2}\partial_{r}\left(\rho r^4\overline{\Omega}U_{2}\right)+\frac{1}{r^2}\partial_{r}\left(\rho\nu_{v}r^4\partial_{r}\overline{\Omega}\right) .
   \label{mean-AM}
  \end{equation}
  Note that only the $l=2$ component of the circulation is able to advect a net amount of angular momentum; the higher order components of $\vec{\mathcal{U}}_{M}$ (for instance those induced in its tachocline by a differentially rotating convection zone) do not contribute to the vertical transport of angular momentum, as was explained in Spiegel  and Zahn (1992).

  \subsection{Horizontal transport of angular momentum}
  
We establish the equation governing the  horizontal transport of angular momentum by  multiplying eq. (\ref{mean-AM}) through $\sin^{2}\theta$ and subtracting it from the original form (\ref{AM-diff-adv}):   \begin{equation}
\rho {{\rm d} \over {\rm d}t}  \left(r^2\sin^2\theta \, \widehat{\Omega}\right)+\vec\nabla\cdot\left(\rho r^{2}\sin^2\theta\, \overline{\Omega} \, \vec{\mathcal{U}}_{M}\right) +
\frac{\sin^2 \theta}{5r^2}\partial_{r}\left(\rho r^4\overline{\Omega} \, U_{2}\right)  = \frac{\sin^2\theta}{r^2}\partial_{r}\left(\rho\nu_{v}r^4\partial_{r}\widehat{\Omega}\right)+\frac{1}{\sin\theta}\partial_{\theta}\left(\rho\nu_{h}\sin^3\theta \, \partial_{\theta}\widehat{\Omega}\right) ;
  \end{equation}
in the advection term we have again neglected the fluctuation $\widehat{\Omega}$ compared to the mean $\overline{\Omega}$. 

The next step is to replace $\widehat{\Omega}(r, \theta)$ by its expansion (\ref{omega-expand}) in the horizontal functions $Q_l(\theta)$. For $l=2$, the equation separates neatly into
  \begin{equation}
\rho {{\rm d} \over {\rm d}t} \left(r^2 \Omega_{2}\right)-2\rho\overline{\Omega}r\left[2V_{2}(r)-\alpha(r)U_{2}\right]=\frac{1}{r^2}\partial_{r}\left(\rho\nu_{v}r^4\partial_{r}\Omega_{2}\right)-10\rho\nu_{h}\Omega_{2} ,
  \end{equation}
    \begin{equation} 
\hbox{with} \quad V_{2}=\frac{1}{6\rho r}\frac{{\rm d}}{{\rm d}r}\left(\rho r^2 U_{2}\right)
\quad  \hbox{ and } \quad 
  \alpha=\frac{1}{2}\frac{{\rm d}\ln\left(r^2\overline{\Omega}\right)}{{\rm d}\ln r} , \nonumber
  \end{equation}   
which we can simplify by assuming that the turbulent transport is much more efficient in the horizontal than in the vertical direction,  i.e.  $\nu_{v} \ll \nu_{h}$:
    \begin{equation}
\rho {{\rm d} \over {\rm d}t} \left(r^2 \Omega_{2}\right)-2\rho\overline{\Omega}r\left[2V_{2}-\alpha U_{2}\right]=-10\rho\nu_{h}\Omega_{2} .
\label{omega2}
  \end{equation}
  In the asymptotic regime  $t \gg  r^2/\nu_{h}$, we retrieve the equation for $\Omega_{2}$ given by Zahn (1992):  
  \begin{equation}
  \nu_{h}\Omega_{2}(r)=\frac{1}{5}r\left[2V_{2}(r)-\alpha(r)U_{2}(r)\right]\overline{\Omega}(r) .
  \end{equation}
  
For $l>2$ the situation is more intricate, with couplings between terms of different $l$.  However, when making the three following assumptions: (i) stationarity ($t \gg  r^2/\nu_{h}$), (ii) strong anisotropy of the turbulent diffusion ($\nu_{v} \ll \nu_{h}$), (iii) vertical advection negligible compared to the horizontal one (thin layer approximation), the $l$-components separate, and one recovers the result given by Spiegel and Zahn (1992) in their treatment of the solar tachocline:
   \begin{equation}
\nu_{h} \Omega_{l}(r)= \overline{\Omega} rV_{l}(r)  ,
   \end{equation}
  namely that horizontal diffusion balances horizontal advection.

  \section{Transport of chemicals}
  
  The transport of chemicals species is governed by an advection/diffusion equation similar to that for the transport of angular momentum: 
 \begin{equation}
\rho  \partial_{t}c_{i}+     \rho\, \vec{\mathcal{U}} \cdot \vec\nabla c_{i} =\frac{1}{r^2}\partial_{r}\left(r^2\rho D_{v}\partial_{r}c_{i}\right)+\frac{1}{r^2\sin\theta}\partial_{\theta}\left(\rho D_{h}\sin\theta\, \partial_{\theta}c_{i}\right) ,
\label{chem-total}
  \end{equation}
  where $c_i$ is the concentration of a given element, and where we assume again that the turbulent diffusivity is anisotropic. For simplicity, we have not included here the term representing the production or destruction of the species due to nuclear reactions, which can be easily added. We follow the same procedure as for the angular momentum: we split the concentration in its horizontal average and its fluctuation 
  \begin{equation}  
  c_{i}(r, \theta)=\overline{c_{i}}(r)+c_{i}'(r,\theta) ,
  \end{equation}
  and  explicit the velocity field in
  \begin{equation}  
 \vec{\mathcal{U}} =  {\vec e}_r \dot{r}  +  \vec{\mathcal{U}}_M
 + {\vec e}_r U_i^{\rm diff}
 \label{vel-field}
  \end{equation}
  where, to the radial displacement accompanying the evolution of the star and to the meridional circulation, we add the microscopic diffusion velocity $U_i^{\rm diff}$, which depends on the considered species. We insert these expressions into  (\ref{chem-total}), which leads us to
   \begin{equation}
\rho {{\rm d} \over {\rm d}t} \left( \overline{c_{i}}+{c_{i}'}\right)+\rho (\mathcal{U}_{M,r}+U_i^{\rm diff}) 
    \partial_{r}\overline{c_{i}} +\rho\,\vec {\mathcal{U}_M} \cdot\vec\nabla{c_{i}'} 
  = \frac{1}{r^2}\partial_{r}\left(r^2\rho D_{v}\partial_r \left({\overline{c_{i}}} + {c_{i}'}\right)\right)+\frac{1}{r^2\sin\theta}\partial_{\theta}\left(\rho D_{h}\sin\theta \,\partial_{\theta}{c}_{i}'\right) .
  \label{total-c}
  \end{equation}  
 Then we  take its horizontal average, making use of the anelastic continuity equation
 $\vec\nabla \cdot (\rho \, \vec{\mathcal{U}}_M) = 0$; we thus obtain the equation governing the evolution of the  mean concentration $\overline{c_{i}}$:  
  \begin{equation}  \rho {{\rm d} \over {\rm d} t } \overline{c_{i}}+\frac{1}{r^2}\partial_{r}\left[r^2\rho<{c_{i}'}\, \mathcal{U}_{M, r}>\right]
  +\frac{1}{r^2}\partial_{r}\left[r^2\rho \overline{c_{i}} U_i^{\rm diff} \right]
  =\frac{1}{r^2}\partial_{r}\left(r^2\rho D_{v}\partial_{r}\overline{c_{i}}\right) ,
\quad \hbox{with} \quad  <f> = {\int_0^\pi f \sin \theta \, {\rm d} \theta \over \int_0^\pi \sin \theta \, {\rm d} \theta} .
\label{mean-c}
  \end{equation}
  
The subtraction of (\ref{mean-c}) from (\ref{total-c}) yields the following equation for the fluctuations of the concentration: 
  \begin{equation}
\rho  {{\rm d} \over {\rm d}t}  {c_{i}'}+\rho\, \mathcal{U}_{M, r}\partial_{r}\overline{c_i}+
\rho\,\vec {\mathcal{U}_M} \cdot\vec\nabla{c_{i}'}-\frac{1}{r^2}\partial_{r} 
\left[r^2\rho<{c_{i}'}\mathcal{U} _{M, r}>\right]=
\frac{1}{r^2}\partial_{r}\left(r^2\rho D_{v}\partial_{r}{c_{i}'}\right)+\frac{1}{r^2\sin\theta}\partial_{\theta}\left(\rho D_{h}\sin\theta\partial_{\theta}{c_{i}'}\right) .
  \end{equation}  
Expanding $c'_i$ and $ \mathcal{U}_{M, r}$ in Legendre polynomials:  
  \begin{equation}
  c_{i}'=\sum_{l>0}\widetilde{c_{i}}_{,l}(r)P_{l}(\cos\theta) \quad \hbox{and}  \quad
  \mathcal{U}_{M, r}=\sum_{l>0}U_{l}(r)P_{l}(\cos\theta) ,
  \end{equation}
and assuming that the gradient of the concentration fluctuations is negligible compared to that of the mean concentration, i.e. $|\vec\nabla c'_i| \ll \partial_r \overline{c_i}$, 
this equation can be recast into  
 \begin{equation}
  \rho\ {{\rm d} \over {\rm d}t} \widetilde{c_{i}}_{,l}+\rho U_{l}\partial_{r}\overline{c_i}=\frac{1}{r^2}\partial_{r}\left(r^2\rho D_{v}\partial_{r}\widetilde{c_{i}}_{,l}\right)-\frac{l(l+1)}{r^2}\rho D_{h}\widetilde{c_{i}}_{,l} .
  \end{equation}
  
As in Chaboyer and Zahn (1992), we further assume that the horizontal diffusivity is much stronger than the vertical one, more precisely that $D_{h} \gg D_{v} ({l_h}/{l_v})^2$,
    where $l_{h} (l_{v})$ is the distance over which $c_{i}'$ changes significantly in the horizontal (vertical) direction, and we finally obtain:  
  \begin{equation}
 {{\rm d} \over {\rm d}t} \widetilde{c_{i}}_{,l}+U_{l}\partial_{r}\overline{c_i}=-\frac{l(l+1)}{r^2}D_{h}\widetilde{c_{i}}_{,l} .
\label{c-tilde}
  \end{equation}
  For $t \gg r^2/D_{h}$ there is an asymptotic state  where
  \begin{equation}
  \widetilde{c_{i}}_{,l}=-\frac{r^2}{l(l+1)D_{h}}U_{l}\partial_{r}\overline{c_i}
  \end{equation}
  and then
  \begin{equation}
  \rho  {{\rm d} \over {\rm d}t} \overline{c_i} +  \frac{1}{r^2}\partial_{r}\left[r^2\rho \overline{c_{i}} U_i^{\rm diff} \right] = \frac{1}{r^2}\partial_{r}\left[r^2\rho (D_{v}+D_{\rm eff}) \partial_{r}\overline{c_i}\right] \quad \hbox{ with } \quad D_{\rm eff}=\sum_{l>0}\frac{r^2\left(U_{l}\right)^{2}}{l(l+1)(2l+1)D_{h}} .
  \label{mean-c-fin}
  \end{equation} 
  However, in order to better resolve the fast evolution phases, one may prefer to retain the time derivative in (\ref{c-tilde}), and solve the equation for the mean concentration (\ref{mean-c}) in the form: 
   \begin{equation}  \rho {{\rm d} \over {\rm d} t } \overline{c_{i}}+\frac{1}{r^2}\partial_{r}\left[r^2\rho
   \sum_{l>0} { \widetilde{c_{i}}_{,l} U_l \over (2l+1)} \right]
  +\frac{1}{r^2}\partial_{r}\left[r^2\rho \overline{c_{i}} U_i^{\rm diff} \right]
  =\frac{1}{r^2}\partial_{r}\left[r^2\rho D_{v}\partial_{r}\overline{c_{i}}\right] .
  \end{equation}
  
Finally, let us recall how the molecular weight $\mu$ is related to the concentrations $c_i$:
  \begin{equation}
  \frac{1}{\mu}=\sum_{i}\frac{1+Z_{i}}{A_{i}}c_{i}
  \end{equation}  
  where $A_{i}$ is the mass number (protons+neutrons) and $Z_{i}$ the number of electrons. From  (\ref{c-tilde}) we can thus derive the following advection/diffusion equation for the horizontal fluctuation of the molecular weight $\Lambda_{l}=\widetilde{\mu_{l}}/\overline{\mu}$: 
  \begin{equation}
  {{\rm d} \over {\rm d}t}  \Lambda_{l} - {{\rm d}  \ln \overline \mu \over {\rm d}t}  \,  \Lambda_{l} =\frac{U_{l}}{H_{p}}\nabla_{\mu}-\frac{l(l+1)}{r^2}D_{h}\Lambda_{l} ,
  \label{lamba-evol}
  \end{equation}
  where we have introduced the pressure scale-height $H_{p}=|{\rm d} r/{\rm d} \ln P |$ and the logarithmic gradient $\nabla_{\mu}={\rm d} \ln\overline{\mu}/ {\rm d}\ln P$. This equation will play a key role in the derivation of the meridional circulation, which we shall discuss in \S6.

  \section{Structural properties of the differentially rotating star}

  \subsection{Baroclinic relation}
  
  In our case where we take a non-uniform and a non-cylindrical rotation law, the centrifugal force does not derive from a potential. We are in the baroclinic configuration, where the isobars and the surfaces of constant density do not coincide anymore. To find how the density varies on an isobar, we start from   the hydrostatic equation:    
  \begin{equation}
  \frac{1}{\rho}\vec\nabla P=\vec g= - \vec\nabla \phi+\vec{\mathcal F}_{\mathcal{C}}\quad \hbox{ with } \quad \vec{\mathcal F}_{\mathcal{C}} =\frac{1}{2}{\Omega}^{2}\vec\nabla\left(r^{2}\sin^{2}\theta\right) 
  \end{equation}
  where $\phi$ is the gravitational potential, and where the local effective gravity  $\vec g$ includes the centrifugal force $\vec{\mathcal F}_{\mathcal{C}}$.   
 Taking the curl of this equation, we get 
  \begin{equation}
  -\frac{1}{{\rho}^{2}}\vec\nabla\rho\wedge\vec\nabla P=-\frac{1}{\rho}\vec\nabla \rho\wedge\vec g=\frac{1}{2}\vec\nabla (\Omega)^{2}\wedge\vec\nabla(r\sin\theta)^{2} ,
  \end{equation}
which to first order takes the form  
 \begin{equation}
-  \frac{\vec\nabla {\rho}^{'}\wedge\vec g}{\overline{\rho}}=\left[\partial_{r}(\Omega^{2})r\cos\theta\sin\theta-\partial_{\theta}(\Omega^{2})\sin^{2}\theta\right]\vec{e}_{\varphi} ,
 \label{barocline}
  \end{equation} 
 with $\rho'$ being the variation of the density on the isobar, which we write as 
  \begin{equation}
\rho^{'}(r, \theta)=\sum_{l>0} \widetilde{\rho_{l}}(r)P_{l}(\cos\theta).
  \end{equation}
 
 Likewise we expand $\Omega^2$ in spherical harmonics; 
 since (cf. \ref{omega-expand}, \ref{fal})
  \begin{equation}
\Omega(r,\theta)=\overline{\Omega}(r)+\sum_{l>0}\Omega_{l}(r)\left(P_{l}(\cos\theta)-I_{l}\right)
\end{equation} 
we have
  \begin{equation}
  \Omega^{2}(r,\theta)=\left[{\overline{\Omega}}^{2}-2\overline{\Omega}\Omega_{2}I_{2}\right]+2\overline{\Omega}\sum_{l>0}\Omega_{l}P_{l}(\cos\theta)
  \label{omega-sq}
  \end{equation}
where we kept only the terms linear in $\Omega_l$.  We expect the largest departures from shellular rotation to occur in the tachoclines, where they are enforced by the differential rotation of the adjacent convection zone; judging by the solar case, such departures should be rather small, of the order of 1/10, thus justifying the linear approximation.

Inserting these expansions in (\ref{barocline}), we reach after some algebra the following expression for the modal amplitudes of the relative density fluctuation on an isobar
    \begin{equation}
\frac{\widetilde{\rho}_{l}(r)}{\overline{\rho}}=\frac{r}{\overline{g}}\mathcal{D}_{l}(r),
\label{baro1}
  \end{equation}
  where $\overline{g}$ is the horizontal average, on the isobar, of the modulus of $\vec g$, and
  \begin{eqnarray}
  \mathcal{D}_{l}(r)&=&\mathcal{N}_{l}^{0}\left\{r\partial_{r}\left[\overline{\Omega}^{2}(r)-2\overline{\Omega}(r)\Omega_{2}(r)I_{2} \right] \frac{1}{3\mathcal{N}_{2}^{0}}\delta_{l,2}\right . \nonumber\\
&+&2r\sum_{s>0}\partial_{r}\left(\overline{\Omega}(r)\Omega_{s}(r)\right)\frac{1}{\mathcal{N}_{s}^{0}}\left[A_{s}^{0}\left(-C_{s-1}^{0}\delta_{l,s-2}+D_{s-1}^{0}\delta_{l,s}\right)+B_{s}^{0}
\left(-C_{s+1}^{0}\delta_{l,s}+D_{s+1}^{0}\delta_{l,s+2}\right)\right]\nonumber\\
  &-&{\left. 2\overline{\Omega}(r)\sum_{s>0}\frac{\Omega_{s}(r)}{\mathcal{N}_{s}^{0}}\left[G_{s}^{0}\left(-C_{s+1}^{0}\delta_{l,s}+D_{s+1}^{0}\delta_{l,s+2}\right)-H_{s}^{0}\left(-C_{s-1}^{0}\delta_{l,s-2}+D_{s-1}^{0}\delta_{l,s}\right)\right]\right\}} .
\label{baro2}
  \end{eqnarray}
All numerical coefficients involved ($A^0_l, B^0_l, C^0_l, D^0_l, G^0_l, H^0_l, \mathcal{N}^0_l$) are given in Appendix A.  We shall display here the result for  $l=2, 4$, where we keep only the first term $\Omega_{2}$ of the expansion of  $\Omega$:  
  \begin{equation}
 \mathcal{D}_{2}
= \frac{1}{3}\left[r\partial_{r}\overline{\Omega}^{2}\right] +\frac{8}{35}\left[r\partial_{r}\left(\overline{\Omega}\Omega_{2}\right)\right]+\frac{8}{7}\overline{\Omega}\Omega_{2}
 \label{barotwo}
  \end{equation}
 \begin{equation}
 \mathcal{D}_{4} = \frac{6}{35} \left[r\partial_{r} \left(\overline{\Omega}\Omega_{2}\right)  
 -2 \overline{\Omega} \Omega_{2}\right] .
 \label{barofour}
  \end{equation}
 For $\Omega_2=0$ we recover the expression given in Zahn (1992): 
 $\widetilde{\rho}_{2}/\overline\rho = (r^2/ 3\overline  g) \partial_r \overline{\Omega}^{2}$.
  
  This baroclinic equation  (\ref{baro1} - \ref{baro2}) plays a key role in linking the density fluctuation on an isobar with the rotation profile. It will allow us to close the system formed by the equation for the transport of angular momentum, that for the transport of the chemical species and that for the transport of heat, which we shall establish in \S6..

  \subsection{Effective gravity} 
  
  Next we examine the redistribution of masses by the centrifugal force and its effect on the gravity inside the star. As all other scalars, we expand the gravity as:   
  \begin{equation}
  g(P,\theta)=\overline{g}(P)+\sum_{l>0}\widetilde{g}_{l}(P)P_{l}(\cos\theta) .
  \end{equation}  
  The goal of this section is to determine the amplitude of the fluctuation on an isobar, $\widetilde{g}_{l}$. The first step will be to calculate the perturbation of the gravitational potential $\phi$ on the sphere of radius $r$, and thus to calculate the  functions $\widehat \phi_{l}(r)$, where we have expanded $\phi$ as:  
  \begin{equation}
  \phi(r, \theta)=\phi^{(0)}(r)+\phi^{(1)}(r,\theta)=\phi_{0}(r)+\sum_{l>0}\widehat{\phi}_{l}(r)P_{l}(\cos\theta)
 \quad \hbox{with} \quad \phi_{0}(r)=-\frac{GM(r)}{r} . 
  \end{equation}  
We follow the method of linearization developed in Sweet (1950), and expand the pressure and the density around the sphere in the same way as $\phi$:   
  \begin{equation}
  P(r, \theta)=P^{(0)}(r)+P^{(1)}(r,\theta)=P_{0}(r)+\sum_{l>0}\widehat{P}_{l}(r)P_{l}(\cos\theta)
  \end{equation}  
  \begin{equation}
  \rho(r, \theta)=\rho^{(0)}(r)+\rho^{(1)}(r,\theta)=\rho_{0}(r)+\sum_{l>0}\widehat{\rho}_{l}(r)P_{l}(\cos\theta) .
  \end{equation}
  
  Then, we take the hydrostatic equation (using the classical definition of a potential, contrary to Zahn 1992):   
  \begin{equation}
  \frac{\vec\nabla P}{\rho}= - \vec\nabla\phi+\vec{\mathcal F}_{\mathcal{C}} \quad \hbox{ where } \quad 
\vec{\mathcal F}_{\mathcal{C}} =\frac{1}{2}\Omega^{2}\vec\nabla(r^2\sin^2\theta) ,
  \end{equation} 
 which we expand to first order: 
   \begin{equation}
  \vec\nabla P^{(0)}= - \rho^{(0)}\vec\nabla\phi^{(0)}
  \end{equation}  
  \begin{equation}
\vec\nabla P^{(1)}= - \rho^{(0)}\vec\nabla\phi^{(1)} - \rho^{(1)}\vec\nabla\phi^{(0)}+
   \rho^{(0)}\vec{\mathcal F}_{\mathcal{C}} .
  \end{equation}   
We eliminate the pressure fluctuation by taking the curl of the latter equation, which gives us:  
  \begin{equation}
  \partial_{\theta}\rho^{(1)} = \frac{\partial_{r}\rho^{(0)}}{g_{0}}\partial_{\theta}\phi^{(1)}+\frac{1}{g_{0}} \left[\rho^{(0)}\partial_{\theta}\left({\mathcal  F}_{{\mathcal C}, r}\right)- 
  \partial_{r}\left(r\rho^{(0)}{\mathcal  F}_{{\mathcal C},\theta}\right)\right] .
  \end{equation}
Next we insert the modal expansion  of $\rho^{(1)}$ and those of the components of the centrifugal force:
\begin{equation}
{\mathcal  F}_{{\mathcal C}, r}(r,\theta)=\sum_{l}a_{l}(r)P_{l}(\cos\theta) \qquad
{\mathcal  F}_{{\mathcal C},\theta}(r,\theta)=-\sum_{l}b_{l}(r)\partial_\theta P_{l}(\cos\theta) ;
\label{ab}
  \end{equation}
 after integration in $\theta$, this yields the modal amplitude  of the density fluctuation over the sphere:
\begin{equation}
\widehat{\rho}_{l}(r) = 
\frac{1}{g_{0}} \left[ \frac{{\rm d}\rho_{0}}{{\rm d}r} \widehat{\phi}_{l} + \rho_{0}a_{l}+\frac{{\rm d}}{{\rm d}r}\left(r\rho_{0}b_{l}\right)\right] .
  \end{equation}  
It remains to insert this expression in the perturbed Poisson equation $\nabla^2 \widehat{\phi}_{l} =
 4 \pi G \widehat{\rho}_{l}$ to retrieve Sweet's result:
  \begin{equation}
  \frac{1}{r}\frac{{\rm d}^2}{{\rm d}r^2}\left(r\widehat{\phi}_{l}\right)-\frac{l(l+1)}{r^2}\widehat{\phi}_{l}-\frac{4\pi G}{g_{0}}\frac{{\rm d}\rho_{0}}{{\rm d}r}\widehat{\phi}_{l}=\frac{4\pi G}{g_{0}}\left[\rho_{0}a_{l}+\frac{{\rm d}}{{\rm d}r}\left(r\rho_{0}b_{l}\right)\right] .
  \label{poisson-pert}
  \end{equation}
The functions $  a_{l}(r)$ and $b_{l}(r)$ are given by 
 \begin{eqnarray}
  a_{l}(r)&=&\frac{2}{3}r\left[\overline{\Omega}^{2}(r)-2\overline{\Omega}(r)\Omega_{2}(r)I_{2}\right]\left(\delta_{l,0}-\delta_{l,2}\right)\label{al}\\ &+& 2r\overline{\Omega}(r)\sum_{s>0}\Omega_{s} (r)\left[\frac{1}{\mathcal{N}_{s}^{0}} \left\{C_{s}^{0} \left(G_{s-1}^{0}\mathcal{N}_{s}^{0}\delta_{l,s}-H_{s-1}^{0}\mathcal{N}_{s-2}^{0}\delta_{l,s-2}\right)
 - D_{s}^{0}\left(G_{s+1}^{0}\mathcal{N}_{s+2}^{0}\delta_{l,s+2}-H_{s+1}^{0}\mathcal{N}_{s}^{0}\delta_{l,s}\right)\right\}\right] \nonumber
  \end{eqnarray} 
 \vskip - 1pc
  \begin{eqnarray}
  b_{l}(r)&=&\frac{1}{3}r\left[\overline{\Omega}^{2}(r)-2\overline{\Omega}(r)\Omega_{2}(r)I_{2}\right]\delta_{l,2} \\
  &+&2r\overline{\Omega}(r)\sum_{s>0}\frac{\Omega_{s}ñ(r)}{\mathcal{N}_{s}^{0}}\left\{A_{s}^{0}\left(-C_{s-1}^{0}\mathcal{N}_{s-2}^{0}\delta_{l,s-2}+D_{s-1}^{0}\mathcal{N}_{s}^{0}\delta_{l,s}\right)
  + B_{s}^{0}\left(-C_{s+1}^{0}\mathcal{N}_{s}^{0}\delta_{l,s}+D_{s+1}^{0}\mathcal{N}_{s+2}^{0}\delta_{l,s+2}\right)\right\} , 
  \nonumber
 \end{eqnarray}
 where we have used the expansion  (\ref{omega-sq})   of $\Omega^2$. The numerical coefficients are given in Appendix A; here we shall quote only the results for $l=2, 4$, with the expansion of $\Omega$ limited to the term $\Omega_2$:
 \begin{eqnarray}
 a_{0}=\frac{2}{3}r \overline{\Omega}^{2} \qquad  \qquad \qquad \qquad 
  b_{0}&=&0  \\ 
a_{2}=-\frac{2}{3} r\overline{\Omega}^{2} + \frac{24}{35}r\overline{\Omega}\Omega_{2} \qquad\;\,
 b_{2}&=&\frac{1}{3}r \overline{\Omega}^{2}+\frac{8}{35}r\overline{\Omega}\Omega_{2} 
 \\
a_{4}=-\frac{24}{35}r\overline{\Omega}\Omega_{2} 
\qquad \qquad\qquad
  b_{4}&=&\frac{6}{35}r\overline{\Omega}\Omega_{2} 
  \end{eqnarray}
 
 With this multipolar expansion of the gravitational potential,  we are ready to calculate the horizontal fluctuation of the effective gravity:  
  \begin{equation}
  \vec g= - \vec\nabla\phi+ \vec{\mathcal F}_{\mathcal{C}}= - \vec\nabla\phi_{0}(r) - \sum_{l}\left[\vec\nabla\widehat{\phi}_{l}(r)P_{l}(\cos\theta)+\widehat{\phi}_{l}(r)\vec\nabla\left(P_{l}(\cos\theta)\right)\right]+\vec{\mathcal F}_{\mathcal{C}}.
  \end{equation}
  
Taking the scalar product of $\vec g$ by itself, and retaining only the first order terms, we get the following expansion for $g^{2}$ over the sphere of radius $r$:  
   \begin{equation}
  g^{2}(r,\theta)=\left[\left(\frac{{\rm d}\phi_{0}(r)}{{\rm d}r}\right)^{2} - 2\frac{{\rm d}\phi_{0}(r)}{{\rm d}r}a_{0}(r)\right]  + 2g_{0}(r)\sum_{l\neq 0}\left(\frac{{\rm d}\widehat{\phi}_{l}(r)}{{\rm d}r} - a_{l}(r)\right)P_{l}(\cos\theta) .
\nonumber
  \end{equation}  
 Comparing with 
   \begin{equation}
  g^2(r,\theta)=\left[g_{0}(r)+\sum_{l\neq 0}\widehat{g}_{l}(r)P_{l}(\cos\theta)\right]^{2}=g_{0}^{2}(r)+2g_{0}(r)\sum_{l\neq 0}\widehat{g}_{l}(r)P_{l}\left(\cos\theta\right) \, ,
  \end{equation}
we identify: 
  \begin{equation}
  \widehat{g}_{l}(r)= \frac{{\rm d}\widehat{\phi}_{l}(r)}{{\rm d} r} - a_{l}(r) .
  \end{equation}
  
Making use of (\ref{xtilde}) we get likewise for the variation along an isobar:  
  \begin{equation}
  \widetilde{g}_{l}(P)=\frac{{\rm d} g_{0}}{{\rm d} r}\frac{1}{g_{0}}\left(\frac{\widehat{P}_{l}}{\rho_{0}}\right) + \frac{{\rm d}\widehat{\phi}_{l}}{{\rm d}r} - a_{l} .
  \end{equation}
 It remains to replace $\widehat{P}_{l}$ by its expression drawn from the  
$\theta$-component of the hydrostatic equation
  \begin{equation}
  \frac{\widehat{P}_{l}}{\rho_{0}}= - \widehat{\phi}_{l}-r b_{l}
  \end{equation}
 to finally obtain the variation of the effective gravity along an isobar: 
  \begin{equation}
  \frac{\widetilde{g}_{l}}{\overline{g}}=-\left[\frac{{\rm d} g_{0}}{{\rm d} r}\frac{1}{{g_{0}^{2}}}rb_{l} + \frac{1}{g_{0}}a_{l}\right] + \frac{{\rm d}}{{\rm d}r}\left(\frac{\widehat{\phi}_{l}}{g_{0}}\right) .
  \end{equation}
  
Keeping only the $\Omega_2$ term and carrying the expansion to $l=4$, we have: 
   \begin{equation}
\frac{\widetilde{g}_{2}}{\overline{g}}=-\frac{{\rm d}g_{0}}{{\rm d}r}\frac{r^2}{{g_{0}^{2}}}\left(\frac{1}{3}\overline{\Omega}^{2} +\frac{8}{35}\overline{\Omega}\Omega_{2}\right)
+ \frac{r}{g_{0}}\left(\frac{2}{3}\overline{\Omega}^{2}  -\frac{24}{35}\overline{\Omega}\Omega_{2}\right) +
  \frac{{\rm d}}{{\rm d}r}\left(\frac{\widehat{\phi}_{2}}{g_{0}}\right)
  \end{equation}  
  \begin{equation}
  \frac{\widetilde{g}_{4}}{\overline{g}}=-\frac{{\rm d}g_{0}}{{\rm d}r}\frac{r^2}{{g_{0}^{2}}}\left(\frac{6}{35}\overline{\Omega}\Omega_{2}\right) + \frac{r}{g_{0}}\left(\frac{24}{35}\overline{\Omega}\Omega_{2}\right) + \frac{{\rm d}}{{\rm d}r}\left(\frac{\widehat{\phi}_{4}}{g_{0}}\right) .
  \label{g-tilde}
  \end{equation}
 For strictly shellular rotation ($\Omega_2=0$) we retrieve the expression of Zahn (1992) (after  changing the sign of $\widehat{\phi}_{2}$): 
   \begin{equation}
\frac{\widetilde{g}_{2}}{\overline{g}}=   {1 \over 3} \overline{\Omega}^{2} \frac{{\rm d}}{{\rm d}r} \left({r^2 \over g_{0}}\right)
    +   \frac{{\rm d}}{{\rm d}r}\left(\frac{\widehat{\phi}_{2}}{g_{0}}\right) .
  \end{equation}
 The two terms represent respectively the contribution of the centrifugal force and that of the modified mass distribution. The latter is obtained by integrating the perturbed Poisson equation (\ref{poisson-pert}) with the appropriate boundary conditions (see \S7).

  \section{Thermal imbalance and transport of heat}
 Following closely the procedure described in Zahn (1992), we now undertake to link the meridional circulation with the rotation profile and the inhomogeneities of chemical composition. 
  We start from the equation stating the conservation of thermal energy:
  \begin{equation}
  \rho T \left[ {\partial S \over \partial t} + \vec{\mathcal{U}} \cdot \vec\nabla{S} \right] =\vec\nabla\cdot\left(\chi\vec\nabla T\right)+\rho\epsilon-\vec\nabla\cdot{\vec{F}}_{h}
 \label{entropy-orig}
  \end{equation} 
  where $S$ is the entropy per unit mass, $\chi$ the thermal conductivity and $\epsilon$ the nuclear energy production rate per unit mass. 
  As Maeder and Zahn (1998), we include the flux ${\vec F}_{h}$  carried by horizontal turbulence. 
  
  We project  this equation on the base of spherical harmonics, starting with the left hand side:
\begin{equation}
 \rho T \left[ {\partial S \over \partial t} + \vec{\mathcal{U}} \cdot \vec\nabla{S} \right] =\overline{\rho}\overline{T}\frac{{\rm d}\overline{S}}{{\rm d} t}+\overline{\rho}\overline{T}\sum_{l>0}
\left[\frac{{\rm d}\widetilde{S}_{l}}{{\rm d}t} + U_l {{\rm d} \overline{S} \over {\rm d}r} \right] P_{l}\left(\cos\theta\right)  ,
  \end{equation} 
where we have used again the expression of the velocity field (\ref{vel-field}) $  \vec{\mathcal{U}} =  \widehat{\vec e}_r \dot{r}  +  \vec{\mathcal{U}}_M $,
together with the modal expansion of the meridional flow (\ref{merid-exp}).
 Recognizing in the horizontal average the production of energy related with the contraction or dilatation of the star during its evolution, i.e.  $\overline{T} {{\rm d}\overline{S}}/{{\rm d}t}=-\epsilon_{grav}$, we rewrite (\ref{entropy-orig}) as
 \begin{equation} 
\overline{\rho}\overline{T}\sum_{l>0}
\left[\frac{{\rm d}\widetilde{S}_{l}}{{\rm d}t} + U_l {{\rm d} \overline{S} \over {\rm d}r} \right] P_{l}\left(\cos\theta\right) =\vec\nabla\cdot\left(\chi\vec\nabla T\right)+\rho\epsilon+\overline{\rho}\overline{\epsilon}_{grav}-\vec\nabla\cdot{\vec F}_{h} ,
\label{entrop-gauche}
  \end{equation}
thus introducing the lagrangian time derivative. Note that by construction the right hand side has now a zero horizontal average.
Then we expand the temperature  in
  \begin{equation}
  T(P,\theta)=\overline{T}(P)+\sum_{l>0}\widetilde{T}_{l}(P)P_{l}(\cos\theta) ,
  \end{equation}
from which we deduce the gradient
  \begin{equation}
  \vec\nabla T=\rho\left[\frac{{\rm d}\overline{T}}{{\rm d}P} +\sum_{l>0}
  \frac{{\rm d}\widetilde{T}_{l}}{{\rm d}P}P_{l}(\cos\theta)\right]\frac{\vec\nabla P}{\rho}+
  \sum_{l>0}\widetilde{T}_{l}\vec\nabla P_{l}(\cos\theta) .
  \end{equation}
Multiplying by the conductivity $\chi$ and taking the divergence, we obtain
\begin{eqnarray}
  \vec\nabla\cdot\left(\chi\vec\nabla T\right)&=&\rho\chi\left[\frac{{\rm d}\overline{T}}{{\rm d}P}+\sum_{l>0}\frac{{\rm d}\widetilde{T}_{l}}{{\rm d}P}P_{l}(\cos\theta)\right]  \vec \nabla \cdot \left( \frac{\vec\nabla P}{\rho}\right) + \vec\nabla\left(\rho\chi\left[\frac{{\rm d}\overline{T}}{{\rm d}P}+\sum_{l>0}\frac{{\rm d}\widetilde{T}_{l}}{{\rm d}P}P_{l}(\cos\theta)\right]\right)\cdot\frac{\vec\nabla P}{\rho}\nonumber\\
  &+&\sum_{l>0}\vec\nabla\left(\chi\widetilde{T}_{l}\right)\cdot\vec\nabla P_{l}(\cos\theta)+\sum_{l>0}\chi\widetilde{T}_{l}\nabla^{2}P_{l}(\cos\theta) .
  \end{eqnarray}  
Next we replace
\begin{equation}
\frac{\vec\nabla P}{\rho}=\vec g= - \vec \nabla \phi +\vec{\mathcal F}_{\mathcal{C}}  \quad \hbox{with} \quad |\vec g| =
\overline{g}(P)  + \sum_{l>0}\widetilde{g}_{l}(P)P_{l}(\cos\theta)    \quad \hbox{and} \quad
\vec{\mathcal F}_{\mathcal{C}} = {1 \over 2} \Omega^2 \vec{\nabla} (r \sin \theta)^2  ,
\label{expansions} 
\end{equation}
and  expand 
\begin{equation}
\rho \chi = \overline{\rho \chi} +\sum_{l>0}\widetilde{\rho \chi}_{l}P_{l}(\cos\theta), \qquad
\nabla^2 \phi = 4 \pi G \overline{\rho} + 4 \pi G \sum_{l>0}\widetilde{\rho}_{l}P_{l}(\cos\theta),   \quad  \vec\nabla\cdot   \vec{\mathcal F}_{\mathcal{C}} 
=\overline{f}_{\mathcal{C}}+\sum_{l>0}\widetilde{f}_{\mathcal{C},l}P_{l}(\cos\theta)  .
\end{equation}
The  functions $\overline{f}_{\mathcal{C}}$ and $\widetilde{f}_{\mathcal{C},l}$ are easily derived. Expressing the components of the centrifugal force in terms of $a_l$ and $b_l$  introduced in (\ref {ab}), we get
\begin{equation}
\vec\nabla\cdot\vec {\mathcal F}_{\mathcal{C}}=\sum_{l=0}^{\infty}\left[\frac{1}{r^2}\partial_{r}\left(r^2 a_{l}(r)\right)+\frac{l(l+1)}{r}b_{l}(r)\right]P_{l}(\cos\theta)
\label{fc-mean}
\end{equation}
and therefore:
\begin{equation}
\overline{f}_{\mathcal{C}}=\frac{1}{r^2}\partial_{r}\left(r^2a_{0}\right) \quad \hbox{and} \;
\widetilde{f}_{\mathcal{C},l}=\frac{1}{r^2}\partial_{r}\left(r^2a_{l}\right)+l(l+1)\frac{b_{l}}{r} .
\label{fc}
\end{equation} 
  
  Then we explicit the turbulent flux; with the assumption that the turbulent diffusivity is much stronger in the horizontal than in the vertical direction ($D_h \gg D_v$), its divergence reduces to
   \begin{equation}
{\vec\nabla}\cdot {\vec F_{h}}=-\frac{1}{r^2\sin\theta} 
\partial_{\theta} \left[\sin\theta D_{h}\rho T \, \partial_{\theta} S(r,\theta)\right]
  \end{equation}
  which we expand in 
 \begin{equation}
  \vec\nabla\cdot\vec F_{h}=\sum_{l>0}\frac{l(l+1)}{r^2}\overline{\rho T}D_{h}\widetilde{S}_{l}P_{l}(\cos\theta) .
  \end{equation}
   
After some arrangements, and keeping only first order terms, this leads us to 
  \begin{eqnarray}
  \lefteqn{\vec\nabla\cdot\left(\chi\vec\nabla T\right)+\rho\epsilon+\overline{\rho}\overline{\epsilon}_{g} - \vec{\nabla} \cdot \vec{F}_h =<\overline{\rho\chi}\frac{{\rm d}\overline{T}}{{\rm d}P}
  \left(-4\pi G\overline{\rho}+\overline{f}_{\mathcal{C}}\right)+\overline{\rho}
  \frac{{\rm d}}{{\rm d}P}\left(\overline{\rho\chi}
  \frac{{\rm d}\overline{T}}{{\rm d}P}\right)\overline{g}^{2}+\overline{\rho}
  \left(\overline{\epsilon}+\overline{\epsilon}_{g}\right)>}  \label{sec-membre} \\  &+&\sum_{l>0}\left\{\left(\overline{\rho\chi}\frac{{\rm d}\widetilde{T}_{l}}{{\rm d}P} +\widetilde{\rho\chi}_{,l}\frac{{\rm d}\overline{T}}{{\rm d}P}\right) \left(-4\pi G\overline{\rho} +\overline{f}_{\mathcal{C}}\right)+\overline{\rho\chi}\frac{{\rm d}\overline{T}}{{\rm d}P} 
  \left(-4\pi G\widetilde{\rho}_{l}+\widetilde{f}_{\mathcal{C},l}\right)
  +\overline{\rho}\frac{{\rm d}}{{\rm d}P}\left(\overline{\rho\chi}
  \frac{{\rm d}\overline{T}}{{\rm d}P}\right)2\overline{g}\widetilde{g}_{l}\right.\nonumber\\
  &+&{\left. \overline{\rho}\frac{{\rm d}}{{\rm d}P}\left(\widetilde{\rho\chi}_{,l}
  \frac{{\rm d}\overline{T}}{{\rm d}P}+\overline{\rho\chi}
  \frac{{\rm d}\widetilde{T}_{l}}{{\rm d}P}\right)\overline{g}^{2} -{l(l+1) \over r^2} \overline{\chi}\widetilde{T}_{l}+\widetilde{\rho}_{l} \frac{{\rm d}}{{\rm d}P}\left(\overline{\rho\chi}
  \frac{{\rm d}\overline{T}}{{\rm d}P}\right)\overline{g}^{2}+\widetilde{\rho\epsilon}_{,l} 
  - {l(l+1) \over r^2} \overline{\rho T} D_h \widetilde{S}_l \right\}}P_{l}(\cos\theta) . \nonumber
  \end{eqnarray}  
  As mentioned above, the horizontal average $<...>$ is zero on a level surface, which means that the star is in average radiative equilibrium,  including the production of nuclear and gravitational energy. 
  
It remains to replace all tilded quantities in terms of two only  (for each $l$-component). Here we choose these two variables to be the horizontal variations of temperature and molecular weight, instead of Zahn (1992) and Maeder \& Zahn (1998) who took the density variation. We define
 \begin{equation}
 \Psi_l = {\widetilde{T}_l \over \overline{T}}  \, , \qquad   \Lambda_l = {\widetilde{\mu}_l \over \overline{\mu}} \, ,
 \qquad \Theta_l= {\widetilde{\rho}_l \over \overline{\rho}} \, ,
   \end{equation}
  and use the equation of state in differential form
  \begin{equation} 
 {{\rm d}\rho \over \rho} = \alpha {{\rm d}P \over P} - \delta {{\rm d}T \over T} + \varphi {{\rm d} \mu \over \mu} 
   \end{equation}
with the straightforward definitions
$ \delta = - ({\partial \ln \rho / \partial \ln T})_{P, \mu}$ etc.,   to express
    \begin{equation}
\Theta_l = \varphi \Lambda_l - \delta \Psi_l .
   \end{equation}
   We operate likewise with the other variables:
 \begin{equation}
  \frac{\widetilde{\chi}_{l}}{\chi}=\chi_{T}\Psi_{l}+\chi_{\mu}\Lambda_{l}, \qquad
   \frac{\widetilde{\epsilon}_{l}}{\epsilon}=\epsilon_{T}\Psi_{l}+\epsilon_{\mu}\Lambda_{l} ,
  \end{equation}
and introduce
 \begin{equation}
 K = {\overline{\chi} \over {\overline{\rho} C_P}}, \qquad
\rho_m = {M(P) \over {4 \over 3} \pi r^3}, \qquad  \epsilon_m   = {L(P) \over M(P)} \quad \hbox{and}  \quad
  f_{\epsilon}=\frac{\overline\epsilon}{\left(\overline{\epsilon}+\overline{\epsilon}_{grav}\right)}  .
  \end{equation}
  Switching to $r$ as independent variable, with ${\rm d}P = - \overline{\rho g} \, {\rm d}r$, we cast  (\ref{sec-membre}) in its final form:   
   \begin{equation}
  \frac{\vec\nabla\cdot\left(\chi\vec\nabla T\right)+\rho\epsilon+\overline{\rho}\overline{\epsilon}_{grav}-\vec{\nabla} \cdot \vec{F}_h}{\overline{\rho}}=\sum_{l>0}\frac{L}{M}\mathcal{T}_{l}P_{l}(\cos\theta)
  \end{equation} 
  with 
  \begin{eqnarray}
  \mathcal{T}_{l}&=&2\left[1-\frac{\overline{f}_{\mathcal{C}}}{4\pi G\overline{\rho}}-\frac{\left(\overline{\epsilon}+\overline{\epsilon}_{grav}\right)}{\epsilon_{m}}\right]\frac{\widetilde{g}_{l}}{\overline{g}}+\frac{\widetilde{f}_{\mathcal{C},l}}{4\pi G\overline{\rho}}
    -\frac{\overline{f}_{\mathcal{C}}}{4\pi G\overline{\rho}}\left(-\delta\Psi_{l}+\varphi\Lambda_{l}\right)\nonumber\\
  &-&\frac{\rho_{m}}{\overline{\rho}}{r \over 3} \partial_r \left(-H_T \partial_r\Psi_{l} +(1-\delta+\chi_{T})\Psi_{l}+(\varphi+\chi_{\mu})\Lambda_{l}\right)
  -\frac{\left(\overline{\epsilon}+\overline{\epsilon}_{grav}\right)}{\epsilon_{m}}\left(-H_{T}\partial_r\Psi_{l}+(1-\delta+\chi_{T})\Psi_{l}+(\varphi+\chi_{\mu})\Lambda_{l}\right)\nonumber\\  &+&\frac{\left(\overline{\epsilon}+\overline{\epsilon}_{grav}\right)}{\epsilon_{m}}\left((f_{\epsilon}\epsilon_{T}-f_{\epsilon}\delta)\Psi_{l}+(f_{\epsilon}\epsilon_{\mu}+f_{\epsilon}\varphi)\Lambda_{l}\right)
    - \frac{\left(\overline{\epsilon}+\overline{\epsilon}_{grav}\right)}{\epsilon_{m}}\left(-\delta\Psi_{l}+\varphi\Lambda_{l}\right)-\frac{\rho_{m}}{\overline{\rho}}\frac{l(l+1)H_{T}}{3r}
  \left[\Psi_{l} + \left(\frac{D_{h}}{K}\right){\widetilde{S_{l} } \over C_P}\right] .
  \label{tcal}
  \end{eqnarray}
 Here we have used the property that
  \begin{equation} \overline{\rho}\overline{\chi}\frac{d\overline{T}}{dP}=\frac{\int\!\!\!\int_{S(r)}\overline{\chi}\vec\nabla\overline{T}\cdot d\vec S}{\int\!\!\!\int_{S(r)}\vec g\cdot d\vec S}=\frac{\int\!\!\!\int_{S(r)}\overline{\chi}\vec\nabla\overline{T}\cdot d\vec S}{\int\!\!\!\int\!\!\!\int_{V(r)}\left[-\nabla^{2}\phi+\vec\nabla\cdot\vec{\mathcal F}_{\mathcal{C}}\right]dV}=\frac{L(r)}{4\pi G M(r) - r^2 a_0(r)} .
  \end{equation}
To lowest order $r^2 a_0= (2/3) r^3 \overline \Omega^2$, according to (\ref{al}), which justifies the approximation
  \begin{equation} 
  \overline{\rho}\overline{\chi}\frac{d\overline{T}}{dP}  = \frac{L(r)}{4\pi G M(r)} .
   \end{equation} 
   
  The last step is to put in final form the left hand side of the heat equation  (\ref{entrop-gauche}).   In a medium of varying composition the entropy of mixing must be taken into account (Maeder \& Zahn 1998). In the simplest case, where the stellar material can be approximated by a simple mixture of hydrogen and helium with a fixed abundance of metals, the entropy of mixing can be expressed in terms of the molecular weight only; then we have 
  \begin{equation}
  {\rm d}S=C_{p}\left[\frac{{\rm d}T}{T}-\nabla_{\mathrm ad}\frac{{\rm d}P}{P}+\Phi(P,T,\mu)\frac{{\rm d}\mu}{\mu}\right] \, ,
  \end{equation}  
  where  $\nabla_{\rm ad}$ is the adiabatic gradient and $\Phi$ is a function of the metal mass fraction and of $\mu$, the mean molecular weight
  (see Maeder \& Zahn 1998). Applying this relation to the fluctuation of entropy around an isobar leads to:   
  \begin{equation}
  \widetilde{S}_{l}=C_{p}\left[\Psi_{l}+\Phi\Lambda_{l}\right] .
  \label{stilde}
  \end{equation}
   In the same way, we get for the mean entropy gradient:
  \begin{equation}
  \partial_{r}\overline{S}=\frac{C_{p}}{H_{p}}\left(\nabla_{\rm ad}-\nabla-\Phi\nabla_{\mu}\right)
  \end{equation}
  with: $\nabla={\rm d} \ln T/ {\rm d} \ln P$ and $\nabla_{\mu}={\rm d} \ln\mu /{\rm d} \ln  P$.
  
  We refer back to (\ref{entrop-gauche}) to express the left hand side of the thermal energy equation, 
  and take into account that $\Lambda_l$ obeys a similar advection/diffusion equation (\ref{lamba-evol}); this permits the elimination of ${\rm d} \Lambda_l / {\rm d}t$ on the l.h.s. and leads us to 
  \begin{equation}
  \overline{T} C_{p}\left[{{\rm d} \Psi_{l} \over {\rm d} t} + \Phi  {{\rm d} \ln \overline \mu \over {\rm d}t} \Lambda_l +  \frac{U_{l}(r)}{H_{p}}\left(\nabla_{ad}-\nabla\right)\right]=\frac{L}{M}{\mathcal T}_{l}(r) ,
    \label{heat1}
  \end{equation}
where we cast ${\mathcal T}_{l}(r)$  (\ref{tcal}) in its final form, having replaced $\widetilde{S}_l$  by (\ref{stilde}) on the right hand side:  
  \begin{eqnarray}
  \mathcal{T}_{l}&=&2\left[1-\frac{\overline{f}_{\mathcal{C}}}{4\pi G\overline{\rho}}-\frac{\left(\overline{\epsilon}+\overline{\epsilon}_{grav}\right)}{\epsilon_{m}}\right]\frac{\widetilde{g}_{l}}
  {\overline{g}}+\frac{\widetilde{f}_{\mathcal{C},l}}{4\pi G\overline{\rho}}
    -\frac{\overline{f}_{\mathcal{C}}}{4\pi G\overline{\rho}}
    \left(-\delta\Psi_{l}+\varphi\Lambda_{l}\right)  \label{tcal-final} \\
  &+&\frac{\rho_{m}}{\overline{\rho}}\left[ \frac{r}{3}\partial_{r}\left(H_T \partial_r\Psi_{l}
   -(1-\delta+\chi_{T})\Psi_{l}-(\varphi+\chi_{\mu})\Lambda_{l}\right)
    - \frac{l(l+1)H_{T}}{3r}
  \left(1 + \frac{D_{h}}{K}\right){\Psi_l}\right]  \nonumber  \\
  &+&\frac{\left(\overline{\epsilon}+\overline{\epsilon}_{grav}\right)}{\epsilon_{m}}
 \left\{ \left(H_{T}   \partial_r\Psi_{l} -(1-\delta+\chi_{T})\Psi_{l}-(\varphi+\chi_{\mu})\Lambda_{l}\right)+(f_{\epsilon}\epsilon_{T} - f_{\epsilon}\delta +\delta)\Psi_{l}+(f_{\epsilon}\epsilon_{\mu}+f_{\epsilon}\varphi - \varphi)\Lambda_{l}\right\} \, . \nonumber 
    \end{eqnarray}
We recall that $\overline{f}_{\mathcal{C}} + \sum {\widetilde{f}_{\mathcal{C},l}}P_l(\cos \theta)$ is the divergence of the centrifugal force (cf. eq. \ref{expansions} ); the functions $\overline{f}_{\mathcal{C}} $ and $ {\widetilde{f}_{\mathcal{C},l}}$ are given in (\ref{fc}).

This expression is similar to that derived in Maeder \& Zahn (1998), where $\Theta_2 =  \widetilde{\rho}_2 / \overline {\rho}$ was used as dependent variable instead of $\Psi_l =  \widetilde{T}_l / \overline {T}$ here. However we note an important difference, namely that the l.h.s. involves here the only the thermal part of the subadiabatic gradient $\left(\nabla_{ad}-\nabla\right)$, whereas in Maeder \& Zahn (1998) it was  $\left(\nabla_{ad}-\nabla + (\varphi/ \delta) \nabla_\mu \right)$. Also, here the highest order derivative at the r.h.s. operates only on $\Psi_l$, whereas in Maeder \& Zahn (1998) it applied both on $\Theta_2$ and on $\Lambda_2$. We expect therefore the present form to be less sensitive, numerically, to steep composition gradients.
  
 The temperature fluctuation on an isobar is thus governed by an advection/diffusion equation, from which we can derive  the radial component of the meridional circulation. It  allows us to link the circulation with its causes, namely both the rotation profile and that of chemical composition. The fact  that this equation has been derived for a general law of rotation allows us to treat simultaneously the bulk of the radiation zones and the tachoclines. 
  
  The system of equations is now completed: we have an advection/diffusion equation for the transport of the angular momentum (mean and fluctuating), another for the transport of chemical elements (mean and fluctuating), another for the temperature (mean and fluctuating), and we have the baroclinic relation which allows us to close the system. It remains to specify the boundaries conditions of this system, to be applied on the limits of the radiation zone.
  
  \section{Boundary conditions}
  
 We shall now review the differential equations we have established for the various quantities which enter in the problem, and state for each of them the appropriate boundary conditions, for a given radiation zone.
To be specific, we consider a star with a radiation zone located between a convective core and an upper convection zone, and designate by $r_{b}$ and $r_{t}$ the radius respectively of the base and of the top of that radiative zone. Of course, in a solar-type  main-sequence star we have $r_b=0$, whereas for a massive main sequence star $r_t=R$, $R$ being the radius of the star.
 
 The equation for the mean angular velocity $\overline \Omega$ (\ref {mean-AM}) is of second order in $r$, and therefore it requires two boundary conditions.  They are obtained by evaluating the budget of angular momentum in each adjacent convective zone: 
    \begin{equation}
 { {\rm d} \over {\rm d}t} \left[\int_{0}^{r_{b}}r^{4}\rho \Omega {\rm d}r \right]={1 \over 5} r^{4}\rho\overline{\Omega}U_{2} \quad \hbox{at} \; r=r_{b}  \qquad
{{\rm d} \over {\rm d}t} \left[\int_{r_{t}}^{R}r^{4}\rho \Omega {\rm d}r\right]=-{1 \over 5} r^{4}\rho\overline{\Omega}U_{2}+\mathcal{F}_{\Omega} \quad \hbox{at} \;  r=r_{t} ,
 \label{bc-mean-omega}
  \end{equation}
   where $\mathcal{F}_{\Omega}$ is the angular momentum flux which is lost at the surface by the stellar wind.  (The perturbation $\Omega_2$ obeys an evolution equation which does not include any derivative in $r$, at least with the approximation we made; therefore it needs no boundary condition.)
 
 Likewise, the equation for the mean concentration (\ref {mean-c}) is of second order in $r$, and its two boundary conditions are obtained by calculating the evolution in time of that mean concentration 
 $\overline{c}_i$ in the two adjacent convection zones (cf. Palacios et al. 2001). We thus have respectively for $r_{t}$ and $r_{b}$:  
  \begin{eqnarray}
 { {\rm d} \over {\rm d}t}\left[\overline{c}_{i}\int_{r_{t}}^{R}r^2\rho {\rm d} r\right]&=&r^{2}\rho\left(U^{\rm diff}_i \overline{c}_{i}\right)-r^{2}\rho\left(D_{v}+D_{\rm eff}\right)\partial_{r}\overline{c}_{i}-\dot{M}\overline{c}_{i} \quad \hbox{at}\; r=r_{t} , \nonumber \\
 { {\rm d} \over {\rm d}t}\left[\overline{c}_{i}\int_{0}^{r_{b}}r^{2}\rho  {\rm d}r\right]&=&-r^{2}\rho\left(U^{\rm diff}_i \overline{c}_{i}\right)+r^{2}\rho\left(D_{v}+D_{\rm eff}\right)\partial_{r}\overline{c}_{i}\quad \hbox{at} \; r=r_{b} ,
  \end{eqnarray}  
 where $\dot{M}$ is the rate of mass loss through the wind.
 
Let us examine now the heat equation (\ref{heat1}-\ref{tcal-final}), which is of second order in $r$ for the variable $\Psi_l$. The variable $\Lambda_l$ need not to be considered here, since it is determined through its evolution equation 
(\ref{lamba-evol}). The boundary conditions on $\Psi_l$ are deduced from the baroclinic equation (\ref{baro1}-\ref{baro2}), which involves $\overline \Omega$ and $\Omega_2$, and their first order derivatives.  We must therefore specify all these functions at the boundaries of the radiation zone. For $\overline \Omega$ this is already done (eq. \ref{bc-mean-omega}), and for $\Omega_{l}(r)$ we require the fluctuations to be continuous at the boundaries:
 \begin{equation}
  \Omega_{l}(r)=\Omega_{l,b} \quad \hbox{at} \; r=r_{b},  \qquad
  \Omega_{l}(r)=\Omega_{l,t} \quad \hbox{at} \;  r=r_{t} ,
  \end{equation}  
 thus assuming that there are no stresses between the radiative and the convective zone. 
From  the baroclinic relation (\ref{barocline}) we deduce that the gradient of the angular velocity must also be continuous, and therefore that
 \begin{eqnarray}
   \partial_{r} \overline{\Omega}(r)=  \partial_{r}\overline{\Omega}_{b} \quad \hbox{at} \; r=r_{b} , \qquad \;\,
   \partial_{r} \overline{\Omega}(r)&=&  \partial_{r}\overline{\Omega}_{t} \quad \hbox{at} \; r=r_{t} , \nonumber\\
   \partial_{r} \Omega_{l}(r)=  \partial_{r}\Omega_{l,b} \quad \hbox{at} \; r=r_{b} ,  \qquad
   \partial_{r} \Omega_{l}(r)&=&  \partial_{r}\Omega_{l,t}\quad \hbox{at} \;  r=r_{t} .
  \end{eqnarray}

     Since we do not solve here for the rotation law in the convective regions (a formidable task!), the best we can afford is to apply a horizontal profile which is inspired from helioseismology, for the base of the outer convection zone. Since the rotation rate $\Omega(r, \theta)$ seems to be independent of $r$ in the solar convection zone, we can take
   \begin{equation}
 \partial_{r} \overline{\Omega}(r)= 0 \quad \hbox{at} \; r=r_{t},  \qquad
   \partial_{r} \Omega_{l}(r)=  0 \; \hbox{at} \quad r=r_{t} .    
  \end{equation}     
 For the condition at the boundary of a convective core, we will have to rely on  numerical simulations  (see the recent results of Browning et al. 2004).
  In the case where we stop the expansion of rotation at the $\Omega_2$ term, and where we assume that the vertical gradient of $\Omega$ is zero, we  obtain:
  \begin{eqnarray}
  \varphi\Lambda_{2}-\delta\Psi_{2}&=&\frac{8}{7}\frac{r}{\overline{g}}\overline{\Omega}\Omega_{2} ,
   \nonumber\\
  \varphi\Lambda_{4}-\delta\Psi_{4}&=&-\frac{12}{35}\frac{r}{\overline{g}}\overline{\Omega}\Omega_{2} .
  \end{eqnarray}
   
   Finally, let us state the boundary conditions for the perturbed Poisson equation. Unlike the precedent equations above, this second order equation must be integrated over the whole star. We require regularity at origin, and continuity with a potential field at the surface ($r=R$):
    \begin{equation}
    \widehat{\phi}_l = 0   \; \hbox{at} \; r=0,   \qquad (l+1)   \widehat{\phi}_l  + {{\rm d} \over {\rm d}r}
      \widehat{\phi}_l  = 0  \; \hbox{at} \; r=R.
  \end{equation}         
  
  \section{Conclusion}
  
  The work presented here marks another step in the description of rotational mixing in stellar radiation zones, after the papers by Zahn (1992) and Maeder \& Zahn (1998). When applied to massive stars, which are fast rotators, such mixing yields much better agreement between models and observations, as was discussed by Meynet and Maeder (2000). However numerical problems arise during the advanced stages of evolution, when steep gradients of composition and rotation develop. For this reason,  our main purpose here was to increase the accuracy of the modelization in latitude  by adding higher order terms to the expansion in spherical harmonics, and also to better resolve the phases of rapid evolution by retaining all time derivatives, except those related to the dynamical relaxation. 
  
  This allows us to include the transport occurring in the tachoclines, where the differential rotation imposed by the adjacent convection zone generates an octupolar circulation, with no net advection of angular momentum. Until now, these layers required an ad-hoc treatment, such as performed by Brun et al. (1999) and Brun et al. (2002), who found that the mixing in the solar tachocline contributes indeed to the depletion of lithium, and that it modifies sufficiently the structure of the star to be detected through helioseismology.
  
  From a more technical point of view, we have written here the heat equation (\ref{heat1}-\ref{tcal-final}),
  which determines the meridional circulation, as a relaxation equation for the temperature fluctuation $\Psi_l$. This breaks the apparent symmetry between the circulation driven by the differential rotation profile (hence by  $\Psi_l$) and the counteracting flow (Mestel's $\mu$-current) which is induced by the inhomogeneities in chemical composition ($\Lambda_l$), since now the expression giving the circulation velocity involves only the first derivative in $\Lambda_l$, while keeping the second derivative of $\Psi_l$. A benefit of that choice is that it renders this differential equation less sensitive, numerically, to the steep composition gradients which arise in the course of stellar evolution.
  
  One weakness of the modelization of rotational mixing remains the description of the turbulence which is generated by the differential rotation. Such turbulence is certainly anisotropic, due to the stable stratification, and it tends therefore to smooth the angular velocity and the chemical composition on horizontal surfaces. This problem is addressed in the companion paper (Mathis et al. 2004), where we discuss the prescriptions for the turbulent transport which are presently available.
  
 A major shortcoming of the present modelization of rotational mixing is that it predicts a rapidly spinning solar core, contrary to the findings of helioseismology. This proves that other physical processes contribute to the transport of angular momentum in solar-type stars, which are presently slow rotators because they have been spun down by their wind. Internal gravity waves, which are generated by turbulent convection at the interface with a convective region, are one possible cause. Their effect has been studied by Talon et al. (2002), and is now being introduced in stellar evolution codes (Talon \& Charbonnel 2003). 
  
  Magnetic stresses are another serious candidate for the angular momentum transport, as was pointed out already by Mestel (1953). The alternating dynamo field does not penetrate enough into the solar radiation zone to produce any impact on the rotation law, as was shown by Garaud (1999). But even a weak fossil field could enforce nearly uniform rotation, although the outcome depends sensitively on the poloidal field  which is assumed, and on whether it penetrates into the differentially rotating convection zone, as illustrated by the calculations performed by Ruediger and Kitchanitov (1996, 1997) and by MacGregor and Charbonneau (1999). Therefore the problem must be treated consistently, taking into account the advection of the field by the meridional circulation, itself being modified by the field, and imposing proper boundary conditions. We shall address this problem in two forthcoming papers, one dealing with an axisymmetric field and the second with a non-axisymmetric field, as observed in oblique rotators. 
  
  \begin{acknowledgements}
  We are grateful to C. Charbonnel, A. Maeder, P. Morel, A. Palacios and S. Talon for their careful reading of the manuscript and their helpful remarks. This work was supported in part by the Centre National de la Recherche Scientifique (Programme National de Physique Stellaire). 
     \end{acknowledgements}

\appendix{}

\section{Algebra related to the spherical harmonics}

The  spherical harmonics are defined by:
\begin{equation}
Y_{l}^{m}(\theta,\varphi)=\mathcal{N}_{l}^{m}P_{l}^{|m|}\left(\cos\theta\right)e^{im\varphi}
\end{equation}
with the normalization:
\begin{equation}
\mathcal{N}_{l}^{m}=(-1)^{\frac{\left(m+|m|\right)}{2}}\left[\frac{2l+1}{4\pi}\frac{(l-|m|)!}{(l+|m|)!}\right]^{\frac{1}{2}} .
\end{equation}
We recall that they obey the well-known differential equation:
\begin{equation}
\frac{1}{\sin\theta}\partial_{\theta}\left(\sin\theta\partial_{\theta}Y_{l}^{m}\left(\theta,\varphi\right)\right)+\frac{1}{\sin^{2}\theta}\partial_{\varphi}^{2}Y_{l}^{m}\left(\theta,\varphi\right)=-l\left(l+1\right)Y_{l}^{m}\left(\theta,\varphi\right) .
\label{ode}
\end{equation}

In deriving certain equations in \S5 we used the following recursion relations for $m=0$:
\begin{equation}
\cos\theta Y_{l}^{0}(\theta)=A_{l}^{0}Y_{l-1}^{0}(\theta)+B_{l}^{0}Y_{l+1}^{0}(\theta)
\quad \hbox{where} \;
A_{l}^{0}=\frac{l}{\sqrt{(2l+1)(2l-1)}} \quad \hbox{and} \; B_{l}^{0}=\frac{(l+1)}{\sqrt{(2l+3)(2l+1)}} ,
\label{r1}
\end{equation}
\begin{equation}
\sin\theta Y_{l}^{0}(\theta)=C_{l}^{0}\partial_{\theta}Y_{l-1}^{0}(\theta)-D_{l}^{0}\partial_{\theta}Y_{l+1}^{0}(\theta)
\quad \hbox{where} \; C_{l}^{0}=\frac{1}{\sqrt{(2l+1)(2l-1)}} \quad \hbox{and}  \; D_{l}^{0}=\frac{1}{\sqrt{(2l+3)(2l+1)}} ,
\label{r2}
\end{equation}
\begin{equation}
\cos\theta\partial_{\theta}Y_{l}^{0}(\theta)=E_{l}^{0}\partial_{\theta}Y_{l-1}^{0}(\theta)+F_{l}^{0}\partial_{\theta}Y_{l+1}^{0}(\theta)
\quad \hbox{where} \;
E_{l}^{0}=\frac{l+1}{\sqrt{(2l+1)(2l-1)}} \quad \hbox{and} \; F_{l}^{0}=\frac{l}{\sqrt{(2l+3)(2l+1)}} ,
\label{r3}
\end{equation}
\begin{equation}
\sin\theta\partial_{\theta}Y_{l}^{0}(\theta)=G_{l}^{0}Y_{l+1}^{0}(\theta)-H_{l}^{0}Y_{l-1}^{0}(\theta)
\quad \hbox{where} \;
G_{l}^{0}=\frac{l(l+1)}{\sqrt{(2l+3)(2l+1)}} \quad \hbox{and} \; H_{l}^{0}=\frac{l(l+1)}{\sqrt{(2l+1)(2l-1)}} .
\label{r4}
\end{equation}
The identities (\ref{r2}) and (\ref{r3}) have been deduced from the two others (\ref{r1}--\ref{r4}), with the help of (\ref{ode}). 

\section{System of transport equations for $l=2, 4$}

In this appendix we give the explicit form of the transport equations for $l=2, 4$, with only the first term $\Omega_2$ being kept in the expansion of the rotation rate $\Omega$.

\subsection{Mean equations}

\subsubsection{mean rotation rate}
\begin{equation}
\rho\frac{{\rm d}}{{\rm d}t}\left(r^{2}\overline{\Omega}\right)=\frac{1}{5r^2}\partial_{r}\left(\rho r^{4}\overline{\Omega}U_{2}\right)+\frac{1}{r^2}\partial_{r}\left(r^{4}\rho\nu_{v}\partial_{r}\overline{\Omega}\right)  \qquad \hbox{(eq.  \ref{mean-AM})} ;
\end{equation}

\subsubsection{mean chemical composition}
   \begin{equation}  \rho {{\rm d} \over {\rm d} t } \overline{c_{i}}
  +\frac{1}{r^2}\partial_{r}\left[r^2\rho \overline{c_{i}} U_i^{\rm diff} \right]
  =\frac{1}{r^2}\partial_{r}\left[r^2\rho (D_{v}+D_{\rm eff})\partial_{r}\overline{c_{i}}\right] 
  \qquad \hbox{(eq.  \ref{mean-c-fin})} .
  \end{equation}

\subsection{System for l=2}

\subsubsection{meridional circulation} 
We rewrite (\ref{heat1}-\ref{tcal-final}) in two first order equations as
\begin{equation}
U_{2}=\frac{L}{M \overline{g}}\left(\frac{P}{\overline{\rho}C_{p}\overline{T}}\right)\frac{1}{\nabla_{ad}-\nabla}\mathcal{B}_{2}
\end{equation}
where we have defined:
\begin{eqnarray}
\mathcal{B}_{2}&=&-2\left[1-{\partial_r (r^3 \overline{\Omega}^2) \over 6 \pi G \overline{\rho} r^2} -\frac{\overline{\epsilon}+\overline{\epsilon}_{grav}}{\epsilon_{m}}\right] \left[\frac{r^2}{\overline{g}^{2}}\frac{{\rm d} \overline{g}}{{\rm d}r}\left(\frac{1}{3}\overline{\Omega}^{2}+\frac{8}{35} \overline{\Omega}\Omega_{2}\right)+\frac{r}{\overline{g}}\left(\frac{24}{35}\overline{\Omega}\Omega_{2}-\frac{2}{3}\overline{\Omega}^{2}\right) - {{\rm d} \over {\rm d}r} \left({\widehat \phi_2 \over g_0} \right) \right]   \nonumber \\
&+&
\frac{1}{\pi G\overline{\rho}} \left[\frac{6}{7}\overline{\Omega}\Omega_{2}
+ \frac{r}{3}\left(\frac{18}{35}\Omega_{2}-\overline{\Omega}\right)\partial_{r}\overline{\Omega}
+\frac{6}{35}r\overline{\Omega}\partial_{r}\Omega_{2}\right]   
+\frac{\rho_{m}}{\overline{\rho}}\left[\frac{r}{3}\partial_{r}\mathcal{A}_{2}-\frac{2H_{T}}{r}\left(1+\frac{D_{h}}{K}\right)\Psi_{2}\right]  \\
&+&\frac{\overline{\epsilon}+\overline{\epsilon}_{grav}}{\epsilon_{m}}\left[\mathcal{A}_{2}+\left(f_{\epsilon}\epsilon_{T}-f_{\epsilon}\delta+\delta\right)\Psi_{2}+\left(f_{\epsilon}\epsilon_{\mu}+f_{\epsilon}\varphi-\varphi\right)\Lambda_{2}\right]-\frac{M}{L}C_{p}\overline{T}\left( {{\rm d}\Psi_{2} \over {\rm d}t} + \Phi  {{\rm d} \ln \overline \mu \over {\rm d}t} \Lambda_2 \right) \nonumber
\end{eqnarray}
and:
\begin{equation}
\mathcal{A}_{2}=H_{T}\partial_{r}\Psi_{2} -\left(1-\delta + \chi_{T}\right)\Psi_{2}
-\left(\varphi+ \chi_{\mu}\right)\Lambda_{2} ,
\end{equation}
and where we have replaced the gravity fluctuation by (\ref{g-tilde}),  $\overline{f}_{\mathcal C}$ by
(\ref {fc-mean}) and $\widetilde{f}_{{\mathcal C}ñ,2}$ by (\ref {fc}). 

\subsubsection{baroclinic relation}
From (\ref{baro1}-\ref {barotwo}) we have:
\begin{equation}
\varphi\Lambda_{2}-\delta\Psi_{2}=\frac{r}{\overline{g}}\left[\frac{8}{7}\overline{\Omega}\Omega_{2}+\frac{2}{3}r\left(\overline{\Omega}+\frac{12}{35}\Omega_{2}\right)\partial_{r}\overline{\Omega}+\frac{8}{35}r\overline{\Omega}\partial_{r}\Omega_{2}\right] .
\end{equation}

\subsubsection{horizontal fluctuation of the molecular weight}
We apply (\ref{lamba-evol}) to $l=2$:
\begin{equation}
\frac{d\Lambda_{2}}{dt}-\frac{d\ln\overline{\mu}}{dt}\Lambda_{2}=\frac{U_{2}}{H_{p}}\nabla_{\mu}-\frac{6}{r^2}D_{h}\Lambda_{2} .
\end{equation}

\subsubsection{Poisson equation}
Likewise, for (\ref{poisson-pert}):
 \begin{equation}
  \frac{1}{r}\frac{{\rm d}^2}{{\rm d}r^2}\left(r\widehat{\phi}_{2}\right)-\frac{l(l+1)}{r^2}\widehat{\phi}_{2}-\frac{4\pi G}{g_{0}}\frac{{\rm d}\rho_{0}}{{\rm d}r}\widehat{\phi}_{2}=\frac{4\pi G}{g_{0}}\left[\rho_{0}a_{2}+\frac{{\rm d}}{{\rm d}r}\left(r\rho_{0}b_{2}\right)\right] .
  \end{equation}

\subsection{System for l=4}

\subsubsection{horizontal shear}
\begin{equation}
\rho \frac{{\rm d}\left(r^{2}\Omega_{2}\right)}{{\rm d}t}-2\rho\overline{\Omega}r\left[\frac{1}{3\rho r}\partial_{r}\left(\rho r^{2} U_{2}\right)-\alpha U_{2}\right]=-10\rho\nu_{h}\Omega_{2}  
 \qquad \hbox{(eq.  \ref{omega2})} .
\end{equation}

\subsubsection{meridional circulation}
We proceed as for $l=2$:
\begin{equation}
U_{4}=\frac{L}{M\overline{g}}\left(\frac{P}{\overline{\rho}C_{p}\overline{T}}\right)\frac{1}{\nabla_{ad}-\nabla}\mathcal{B}_{4}
\end{equation}
where we have defined:
\begin{eqnarray}
\mathcal{B}_{4}&=&-2\left[1- {\partial_r (r^3 \overline{\Omega}^2) \over 6 \pi G \overline{\rho} r^2} -\frac{\overline{\epsilon}+\overline{\epsilon}_{grav}}{\epsilon_{m}}\right]
\left[\frac{r^2}{\overline{g}^{2}}\frac{{\rm d}\overline{g}}{{\rm d}r} \left(\frac{6}{35}\overline{\Omega}\Omega_{2}\right)-\frac{r}{\overline{g}}\left(\frac{24}{35}\overline{\Omega}\Omega_{2}\right)- {{\rm d} \over {\rm d}r} \left({\widehat \phi_4 \over g_0} \right)\right]  \nonumber \\
&+&\frac{1}{\pi G \overline{\rho}}\left[\frac{6}{35}\left(2\overline{\Omega}\Omega_{2}-r\left[\Omega_{2}\partial_{r}\overline{\Omega}+\overline{\Omega}\partial_{r}\Omega_{2}\right]\right)\right]+ \frac{\rho_{m}}{\overline{\rho}}\left[\frac{r}{3}\partial_{r}\mathcal{A}_{4}-\frac{20H_{T}}{3r}\left(1+\frac{D_{h}}{K}\right)\Psi_{4}\right] \\
&+&\frac{\overline{\epsilon}+\overline{\epsilon}_{grav}}{\epsilon_{m}}
\left[\mathcal{A}_{4}+\left(f_{\epsilon}\epsilon_{T}-f_{\epsilon}\delta+\delta\right)\Psi_{4}
+\left(f_{\epsilon}\epsilon_{\mu}+f_{\epsilon}\varphi-\varphi\right)\Lambda_{4}\right]
 - \frac{M}{L} C_{p}\overline{T} \left( {{\rm d}\Psi_{4} \over {\rm d}t} + \Phi  {{\rm d} \ln \overline \mu \over {\rm d}t} \Lambda_4 \right) \nonumber
\end{eqnarray}
and:
\begin{equation}
\mathcal{A}_{4}=H_{T}\partial_{r}\Psi_{4}-\left(1-\delta + \chi_{T}\right)\Psi_{4}
-\left(\varphi+ \chi_{\mu}\right)\Lambda_{4}
\end{equation}

\subsubsection{baroclinic relation}
We restate (\ref{baro1}-\ref {barotwo}) as
\begin{equation}
\varphi\Lambda_{4}-\delta\Psi_{4}=\frac{6}{35}\frac{r}{\overline{g}}\left[r\Omega_{2}\partial_{r}\overline{\Omega}+r\overline{\Omega}\partial_{r}\Omega_{2}-2\overline{\Omega}\Omega_{2}\right] .
\end{equation}

\subsubsection{horizontal fluctuation of the molecular weight}
\begin{equation}
\frac{{\rm d}\Lambda_{4}}{{\rm d}t}-\frac{{\rm d}\ln\overline{\mu}}{{\rm d}t}\Lambda_{4}=\frac{U_{4}}{H_{p}}\nabla_{\mu}-\frac{20}{r^2}D_{h}\Lambda_{4}
 \qquad \hbox{(eq. \ref{lamba-evol})} .
\end{equation}

\subsubsection{Poisson equation}
 \begin{equation}
  \frac{1}{r}\frac{{\rm d}^2}{{\rm d}r^2}\left(r\widehat{\phi}_{4}\right)-\frac{l(l+1)}{r^2}\widehat{\phi}_{4}-\frac{4\pi G}{g_{0}}\frac{{\rm d}\rho_{0}}{{\rm d}r}\widehat{\phi}_{4}=\frac{4\pi G}{g_{0}}\left[\rho_{0}a_{4}+\frac{{\rm d}}{{\rm d}r}\left(r\rho_{0}b_{4}\right)\right] 
 \qquad \hbox{(eq.  \ref{poisson-pert})} .
  \end{equation}

These equations are ready to be implemented in a stellar evolution code, together with the boundary conditions discussed in \S7.

  \end{document}